\PassOptionsToPackage{table}{xcolor}
\documentclass[11pt]{article} 

\usepackage{changepage}
\usepackage[justification=centering]{caption}
\usepackage{url}
\usepackage[margin=0.8in]{geometry}

\usepackage{newtxtext}
\usepackage{amsmath, amssymb}  
\usepackage{newtxmath} 

\usepackage{setspace}
\usepackage[tiny,compact]{titlesec}
\usepackage{placeins}
\usepackage[numbers]{natbib}

\usepackage{amsmath, amssymb}
\usepackage[final]{microtype}
\usepackage{xcolor}
\usepackage{float}
\usepackage{textgreek}

\newcommand{\comment}[1]{}

\usepackage{graphicx}
\graphicspath{ {./images} }
\usepackage[utf8]{inputenc}

\usepackage{tikz}
\usetikzlibrary{shapes.geometric, arrows, fit, positioning}

\tikzstyle{mainb} = [rectangle, rounded corners, minimum width=3cm, minimum height=1cm, text centered, draw=black, fill=blue!20, text width=3cm]
\tikzstyle{arrow} = [thick,->,>=stealth]

\usepackage{chronology,geometry}
\tikzset{flippedeventlabel/.append style={align=center}}

\usepackage{makecell}
\usepackage{graphicx}
\usepackage{pdflscape}
\usepackage{longtable}
\usepackage{multirow}
\usepackage{enumitem}
\usepackage{longtable}
\usepackage{array}
\usepackage{amsmath}
\usepackage{hyperref}
\usepackage{tabularx}
\usepackage{adjustbox}
\usepackage{hhline}

\usepackage{tcolorbox}

\begin{document}
\newcommand{\STAB}[1]{\begin{tabular}{@{}c@{}}#1\end{tabular}}

\definecolor{headercolor}{rgb}{0.772, 0.796, 0.847}
\definecolor{altrowcolor}{rgb}{0.882, 0.894, 0.922}
\definecolor{lighteraltrowcolor}{rgb}{0.953, 0.957, 0.965}


\begin{center}
\large{\bf High-Intensity Helical Flow: A Double-Edged Sword in Coronary Artery Haemodynamics}

\large{ C. Shen\textsuperscript{1,*}, M. Zhang\textsuperscript{1}, H Keramati\textsuperscript{1}, D. Almeida\textsuperscript{2}, S. Beier\textsuperscript{1}}

\large{\textsuperscript{1} School of Mechanical and Manufacturing Engineering
University of New South Wales, Sydney NSW 2052, Australia
}

\large{\textsuperscript{2} Virtonomy GmbH, Paul-Heyse-Straße 6, 80336 Munich, Germany}

\large{*Corresponding author}

\large{Email: chi.shen@student.unsw.edu.au}

\end{center}

\newpage
\section*{Abstract}

The role of Helical Flow (HF) in human coronary arteries remains uncertain, yet its understanding promises unprecedented insights into atherosclerotic processes. In this study, we investigated the effects of HF and key haemodynamic descriptors in 39 patient-specific left coronary artery trees of the ASOCA dataset, 20 non-stenosed and 19 stenosed. Absolute HF intensity $h_2$ correlated with higher Time-Averaged Endothelial Shear Stress (TAESS) in all vessel segments regardless of stenoses ($p < 0.05$). In stenosed cases, this correlation was so prominent that the vessel area exposed to adversely low TAESS reduced ($< 0.5$ Pa,$p = 0.0001$), and simultaneously, areas of adversely high TAESS increased ($> 4.71$ Pa, $p < 0.05$) coinciding with high $h_2$ regions. This suggests that HF in coronaries is not always protective as previously thought because it not only mitigates low TAESS associated with long-term plaque development and restenosis, but also exacerbates adversely high TAESS linked to plaque vulnerability increase and even acute events. Our findings redefined the current understanding of the role of helical blood flow in cardiovascular atherosclerotic disease processes.

\newpage
\section{Introduction}

Helical Flow (HF) naturally exists in vascular systems \cite{liu2015_HF_physiology}, characterised by a downstream and rotational motion. Studies on the aorta \cite{liu2009HF_aorta,morbiducci2011_hf_aorta,morbiducci2009_invivo_HF} and carotid bifurcations \cite{gallo2012helical_carotid} have highlighted the favourable effect of HF on haemodynamics by reducing flow disturbances marked by a reduced luminal area exposed to adversely low Endothelial Shear Stress (ESS) and its cardiac-cycle Time Average (TAESS) \cite{stone2007__lowTAWSS}, associated with a reduced risk of plaque progression \cite{Timmins2017_WSS_effect,Baeyens2016WSS_effect} after disease onset \cite{Zhang2024_plaque_onset_and_progression}. Contrary to this, HF intensity is also known to be higher in curved and stenosed segments, which are common arterial regions of adverse clinical events. Specifically, in human-idealised and patient-specific coronaries, severe vessel curvature and torsion were both linked to increased HF intensity (sample size = 3) \cite{Vorobtsova2016tortuosity_BA,Chiastra2017Healthy_and_diseased,Shen2021Secondary_flow}. Severe curvature (often referred to as tortuosity \cite{Kashyap2022_tortuosity_definition}) has been linked to adverse clinical outcomes such as artery stenosis \cite{Tuncay2018invivo_tortuosity_CTA}, Spontaneous Coronary Artery Dissection (SCAD) \cite{Candreva2023_SCAD}, and myocardial ischemia \cite{ZebicMihic2023_NonObstructive_CAD,Estrada2022_tortuosity_MI}. This underscores a contradiction between the reported protective effects of high HF intensity and its vessel-specific occurrence \cite{Tuncay2018invivo_tortuosity_CTA,Candreva2023_SCAD,ZebicMihic2023_NonObstructive_CAD,Estrada2022_tortuosity_MI}. 
Similarly to human arteries, torsion positively correlated to HF intensity in swine arteries \cite{Nisco2020HF}.  Moreover, in swine, high absolute HF helicity $h_2$, correlated to higher ESS magnitudes \cite{Nisco2019HF}. Extremely high ESS has been linked to expansive remodelling \cite{Samady2011}, destabilisation and rupture of plaque \cite{Eshtehardi2017_high_WSS,bajraktari2021_highTAWSS,fukumoto2008_highTAWSS}, thus driving major adverse cardiac events \cite{Kumar2018highwss,Candreva2022highwss}. Additionally, HF  $h_2$ was found to be significantly higher in human coronaries compared to swine \cite{Nisco2021comparison_swine_human}. Therefore, HF studies on high ESS and TAESS are warranted. The relationship between HF and the extreme threshold of these haemodynamic quantities has never been considered in previous human coronary studies. 
Moreover, stenosis was found to significantly increase the helicity intensity in idealised coronary bifurcations \cite{Chiastra2017Healthy_and_diseased}. It is understood that in stenosed arterial regions, plaque continues to grow immediately downstream, adjacent to the existing stenosis \cite{Jahromi2019_disease,Indraratna2022PlaqueProgression}, with high TAESS at the stenosis where the luminal diameter is most narrow and low TAESS downstream immediately adjacent to the stenosis regions \cite{Frattolin2015_disease,Kamangar2017_disease_hyperemic}. An intricate relationship becomes apparent, which warrants further HF-specific studies in patient-specific stenosed human arteries. 
Overall, given the important relationship between HF intensity and ESS/TAESS distributions, several studies have explored how HF is influenced by coronary artery geometries, aiming to establish a link between arterial geometry, HF, and haemodynamics. These HF studies were either based on swine \cite{Nisco2019HF,Nisco2020HF}, which may not directly translate to human coronaries \cite{Nisco2021comparison_swine_human}, non-coronary artery vessels \cite{liu2009HF_aorta,morbiducci2011_hf_aorta,morbiducci2009_invivo_HF,gallo2012helical_carotid}, only considered the left main coronary bifurcations \cite{Chiastra2017Healthy_and_diseased}, idealised/modified vessel geometries \cite{Chiastra2017Healthy_and_diseased,Shen2021Secondary_flow} or a limited number (sample size=3) of patient-specific coronary arteries without vessel-specific analysis \cite{Vorobtsova2016tortuosity_BA}. 
Here, for the first time, we included the adversely high TAESS in the investigation of HF quantified by helicity intensity ($h_1$ and $h_2$) and their balance ($h_3$ and $h_4$). The haemodynamics and coronary geometrical effects were analysed in 39, 20 non-stenosed and 19 stenosed, left coronary trees to deliver a human, non-stenosed versus stenosed, patient-specific, whole trees and large sample size consideration. Besides TAESS, we also considered the Oscillatory Shear Index (OSI) and Relative Residence Time (RRT) for haemodynamics. For the geometrical features, we considered the curvature, torsion and diameter. This will elucidate the conflicting associations between the reported favourable haemodynamic effect of high HF intensity whilst also coinciding with vulnerable arterial tree regions prone to adverse clinical events.

\section{Materials and Methods}
\subsection{Patient-specific Geometries}

The open-source ASOCA dataset contains 40 left coronary artery trees \cite{GharleghiASOCA2022,Gharleghi2023ASOCADATA},  20 non-stenosed and 20 stenosed.  Here, we excluded one case because of extreme stenosis ($>90$\%), whereby the limited spatial imaging resolution would introduce large haemodynamic computation uncertainties in line with other published work \cite{Zhang2024_plaque_onset_and_progression}. The dataset acquisition is described elsewhere \cite{Gharleghi2023ASOCADATA}, briefly the trees were reconstructed from Computed Tomography Coronary Angiography (CTCA) acquired via a GE LightSpeed 64-slice CT scanner with an ECG-gated retrospective acquisition protocol. The in-plane image resolution was 0.3 - 0.4 mm, and the out-of-plane resolution was 0.625 mm. Three experts independently segmented the images to derive a majority agreement for a high fidelity dataset. The distal branches were trimmed at locations where the diameter was smaller than 2 mm due to the limited CTCA resolution. More details are provided in the relevant literature \cite{GharleghiASOCA2022,Gharleghi2023ASOCADATA}.

\subsection{Computational Fluid Dynamics Setup}

The left coronary artery trees were simulated using ANSYS CFX (ANSYS Inc., Canonsburg, PA, USA). The discretisation was performed using ICEM-CFD, embedded in the ANSYS package (version 2023R1, Canonsburg, PA, USA). A mesh sensitivity analysis was conducted before the simulations to ensure the computational models' accuracy, efficiency and reliability. The blood flow was assumed to be incompressible and non-Newtonian using a Carreau-Yasuda fluid model \cite{razavi2011rheological}.  A laminar blood model was used for simulations. The maximum Reynolds number was below 2,000 for all cases at all time steps. Since patient-specific flow conditions were not available, an allometric scaling law was used to scale a standard velocity waveform, allowing for a physiologically relevant approximation \cite{Nichols2022_waveform}, with the volumetric flow rate at the inlet calculated as follows \cite{Giessen2011_boundary_conditions}:

\[Q = 1.43 d^{2.55}\]
where Q is the cycle-averaged flow rate, and d is the mean diameter of the left main branch. For each bifurcation, a flow-split outflow strategy was applied to determine the flow rate in the distal branches \cite{Giessen2011_boundary_conditions}:

\[\frac{Q_{sb}}{Q_{mb}} = \left( \frac{d_{sb}}{d_{mb}} \right)^{2.27}\]
where $Q_sb$ and $Q_mb$ are flow rates, and  $d_sb$ and $d_mb$ are the mean diameter of the side and main branch, respectively. The scaling law and flow split are effective due to their strong fit to \textit{in vivo} data, improved by empirical adjustments to Murray’s Law \cite{Giessen2011_boundary_conditions}. In stenosed arteries, minor haemodynamic differences were observed across varying stenosis degrees under rest conditions when using flow split and a lumped parameter model \cite{Zhang2023outflow,Zhang2024_plaque_onset_and_progression}. The artery wall was assumed to be rigid, and a standard no-slip condition was applied \cite{Eslami2020_wall_properties}. A steady-state simulation was performed for each model, and its results were used as the initial condition for the transient simulations of four consecutive cardiac cycles. The results from the fourth cycle were extracted for analysis to minimise transient start-up effects.

\subsection{Coronary Geometrical and Haemodynamics Descriptors}

\begin{figure}
    \centering
    \includegraphics[width=1\linewidth]{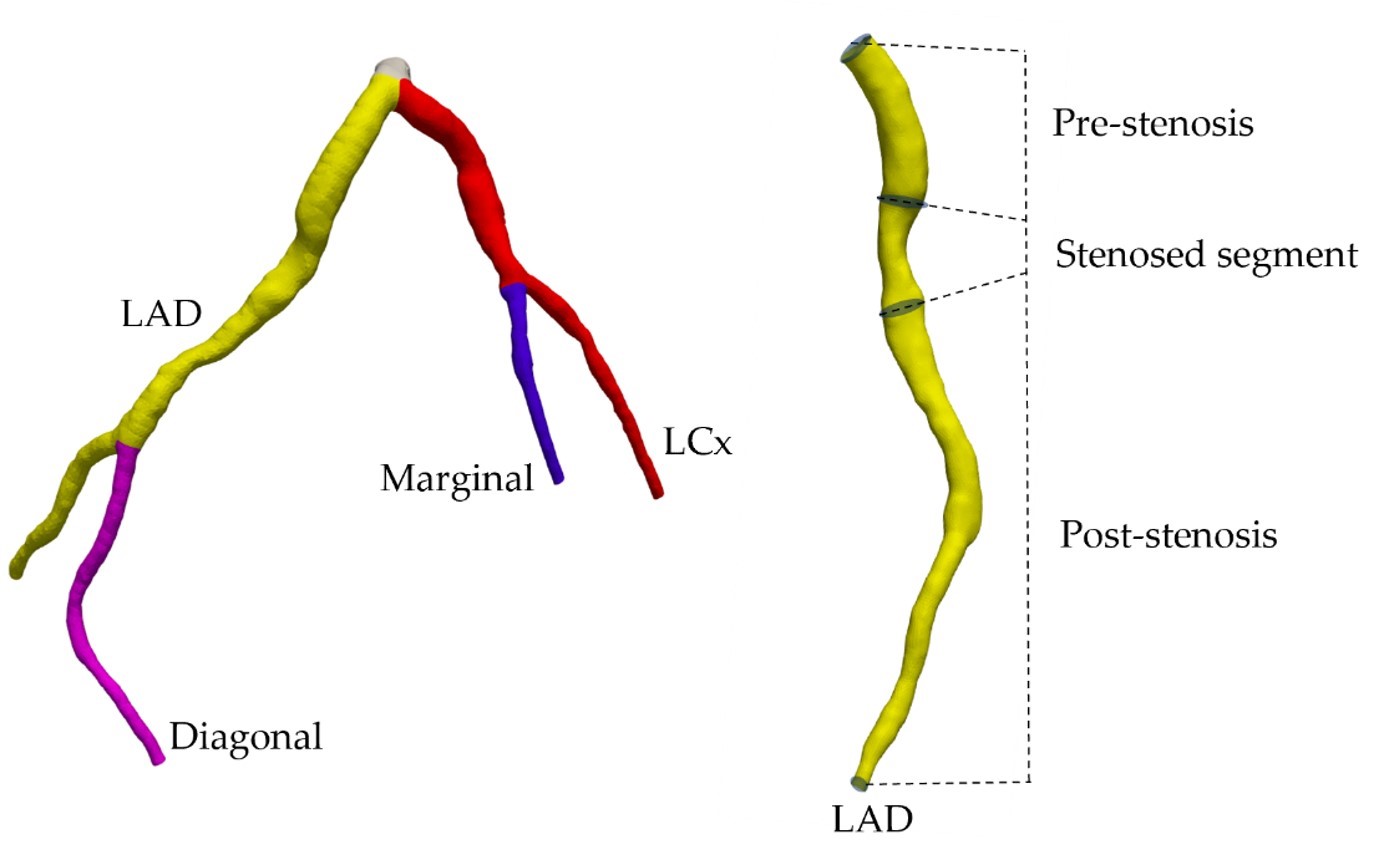}
    \caption{The breakdown of the left coronary tree into Left Anterior Descending LAD (yellow), Left Circumflex artery LCx (red), Diagonal (pink) and Marginal (blue) segments (left). Vessel segments with stenosis were further analysed per pre-stenosis, stenosed, and post-stenosis segments (right). }
    \label{HF_figure1}
\end{figure}

Both 3D geometric and haemodynamic descriptors were extracted from each coronary tree, including the Left Anterior Descending (LAD), Left Circumflex (LCx), Diagonal, and Marginal arteries (Figure \ref{HF_figure1} left). The stenosed regions were defined as the point of minimum luminal diameter extending two vessel diameters upstream and downstream \cite{Scarsini2022_Diffuse_Focal_and_Serial_Lesions} (Figure \ref{HF_figure1} right). The stenosed vessel segments were further subdivided into pre-stenosis, stenosed, and post-stenosis regions for a detailed analysis of the stenotic effects (Figure \ref{HF_figure1} right).
The geometrical descriptors considered include average absolute curvature (commonly referred to as tortuosity \cite{Kashyap2022_tortuosity_definition}), diameter, and torsion, which were automatically calculated using in-house codes \cite{Pau2017shape_in_normal_population,pau2016atlas} using the Vascular Modelling ToolKit (VMTK) :

\[
\kappa_s = \frac{1}{L} \int_{s_1}^{s_2} \frac{\left| \mathbf{c}'(s) \times \mathbf{c}''(s) \right|}{\left| \mathbf{c}'(s) \right|^3} \, ds
\]
where $\mathbf{c}'(s)$ and $\mathbf{c}''(s)$ are the first and second derivatives of the curve $\mathbf{c}$, and $L$ is the length of the curve. The torsion $\tau_a$ was calculated based on:
\[
\tau_a = \frac{1}{L} \int_{s_1}^{s_2} \frac{\left| \left( \mathbf{c}'(s) \times \mathbf{c}''(s) \right) \cdot \mathbf{c}'''(s) \right|}{\left| \mathbf{c}'(s) \times \mathbf{c}''(s) \right|^2} \, ds
\]
where $\mathbf{c}'''(s)$ is the third derivative of the curve $\mathbf{c}$.

The HF was visualised using the Local Normalised Helicity (LNH) iso-surfaces, where right-handed and left-handed rotational flow patterns were colour-coded with red and blue, respectively. HF-based descriptors were used to quantify the density and rotational directions of HF \cite{Nisco2019HF,Nisco2020HF}. Specifically, the cycle-average helicity ($h_1$) and absolute helicity intensity ($h_2$) were calculated, along with signed ($h_3$) and unsigned ($h_4$) helical rotation balance.
Adverse ESS distribution relates to endothelial dysfunction, promoting disease development \cite{Timmins2017_WSS_effect,Baeyens2016WSS_effect}. Low ESS contributes to plaque progression and constrictive arterial wall remodelling \cite{stone2007__lowTAWSS}. Compared to low ESS, the area exposed to high ESS is related to expansive remodelling, increased plaque vulnerability with a higher necrotic core area and higher plaque burden \cite{Eshtehardi2017_high_WSS,bajraktari2021_highTAWSS}, which can further lead to plaque rupture \cite{fukumoto2008_highTAWSS} and acute cardiac events \cite{Kumar2018highwss}. In this work, we analysed the cardiac cycle averaged ESS (TAESS) to account for shear stress variations over the entire cardiac cycle, providing a more comprehensive measure to understand long-term effects. Several factors have been derived from TAESS. OSI quantifies the oscillation in the shear force direction at the vessel wall, where high OSI is associated with lipid accumulation and plaque erosion \cite{Adriaenssens2021PlaqueProgression}. RRT indicates the regions with low TAESS exposure and high particle residence time. High RRT is related to atherosclerotic plaque calcification and necrosis \cite{Kok2019multidirectional_plaqueprogression}. Commonly used thresholds in previous literature were adopted to define adverse values for TAESS (low $< 0.5$ Pa \cite{BeierBifurcation2016} and high $> 4.71$ Pa \cite{Kumar2018highwss,Candreva2022highwss}), high OSI ($>0.1$ \cite{Xie2014tortuosity}), and high RRT ($> 4.17 \, \text{Pa}^{-1}$ \cite{Kok2019multidirectional_plaqueprogression}). For each vessel segment, TAESS was calculated as the average time-averaged endothelial shear stress over the vessel lumen area. We also reported HF-based descriptors ($h_1$, $h_2$, $h_3$ and $h_4$), TAESS and the normalised percentage of vessel area exposed to the adverse haemodynamics thresholds. All considered descriptors are shown in Table 1.

\renewcommand{\arraystretch}{1.5} 

\begin{table}[ht]
\centering
\caption{Definitions of helical flow and haemodynamic descriptors.}
\begin{tabularx}{\textwidth}{|l|X|X|}
\hline
\textbf{Descriptors} & \textbf{Equations} & \textbf{Definitions} \\ \hline
\multicolumn{3}{|c|}{\textbf{Helical flow-based descriptors}} \\ \hline
$h_1$ & 
$\frac{1}{T V} \int_{0}^{T} \int_{V} \mathbf{v}(\mathbf{x},t) \cdot \boldsymbol{\omega}(\mathbf{x},t) \, dV \, dt$ & 
Signed average helicity \\ \hline
$h_2$ & 
$\frac{1}{T V} \int_{0}^{T} \int_{V} |\mathbf{v}(\mathbf{x},t) \cdot \boldsymbol{\omega}(\mathbf{x},t)| \, dV \, dt$ & 
Average helicity intensity \\ \hline
$h_3$ & 
$\frac{h_1}{h_2}$ & 
Signed balance of HF rotations \\ \hline
$h_4$ & 
$\frac{|h_1|}{h_2}$ & 
Unsigned balance of HF rotations \\ \hline
LNH & 
$\frac{\mathbf{v}(\mathbf{x},t) \cdot \boldsymbol{\omega}(\mathbf{x},t)}{|\mathbf{v}(\mathbf{x},t)||\boldsymbol{\omega}(\mathbf{x},t)|}$ & 
Visualisation of the helical flow \\ \hline
\multicolumn{3}{|c|}{\textbf{ESS-based haemodynamic descriptors}} \\ \hline
ESS & 
$\mathbf{n} \cdot \boldsymbol{\tau}_{ij}$ & 
Shear stress along the vessel wall \\ \hline
TAESS & 
$\frac{1}{T} \int_{0}^{T} |\mathbf{n} \cdot \boldsymbol{\tau}_{ij}| \, dt$ & 
The cardiac cycle averaged ESS \\ \hline
OSI & 
$\frac{1}{2} \left( 1 - \frac{\int_{0}^{T} |\tau_{w}| \, dt}{\int_{0}^{T} \tau_{w} \, dt} \right)$ & 
Representation of the change of the ESS vector from a predominant flow direction \\ \hline
RRT & 
$\frac{1}{(1 - 2 \, \text{OSI}) \, \text{TAESS}}$ & 
The residence time of elements in the blood adjacent to the wall \\ \hline
\end{tabularx}
\label{tab:helical_flow}
\vspace{1ex} 
\raggedright
\textbf{Note:} $T$: cycle period, $V$: volume of interest, $\mathbf{v}(\mathbf{x},t)$: velocity field at the location $\mathbf{x}$ and time $t$, $\boldsymbol{\omega}(\mathbf{x},t)$: vorticity, $\boldsymbol{\tau}_{ij}$: endothelial shear stress tensor, $\mathbf{n}$: normal vector at the vessel wall, $\tau_w$: shear force at the vessel wall.
\end{table}

\subsection{Statistical Analysis}

Statistical analyses were performed using the Python Statsmodels package, version 0.14.2. The normality of distribution was assessed with the Shapiro–Wilk test. Continuous variables with a normal distribution are presented as mean ± Standard Deviation (SD), while non-normally distributed variables are presented as median and inter-quartile range [IQR]. For comparisons between non-stenosed and stenosed groups, Welch’s t-tests were used for normally distributed variables and the Mann-Whitney U-test for non-normally distributed variables. Spearman correlation coefficients, ρ, and the regression p-value was used to evaluate the correlations between HF-based and haemodynamics, geometrical and HF-based, and geometrical and haemodynamic descriptors. To avoid type-I error in the statistical hypothesis testing, p-values presented in this work were adjusted with the Bonferroni-Holm correction, with a p-value after adjustment $< 0.05$ considered statistically significant.

\section{Results}
\subsection{Comparisons between non-stenosed and stenosed groups }
\begin{figure}
    \centering
    \includegraphics[width=1\linewidth]{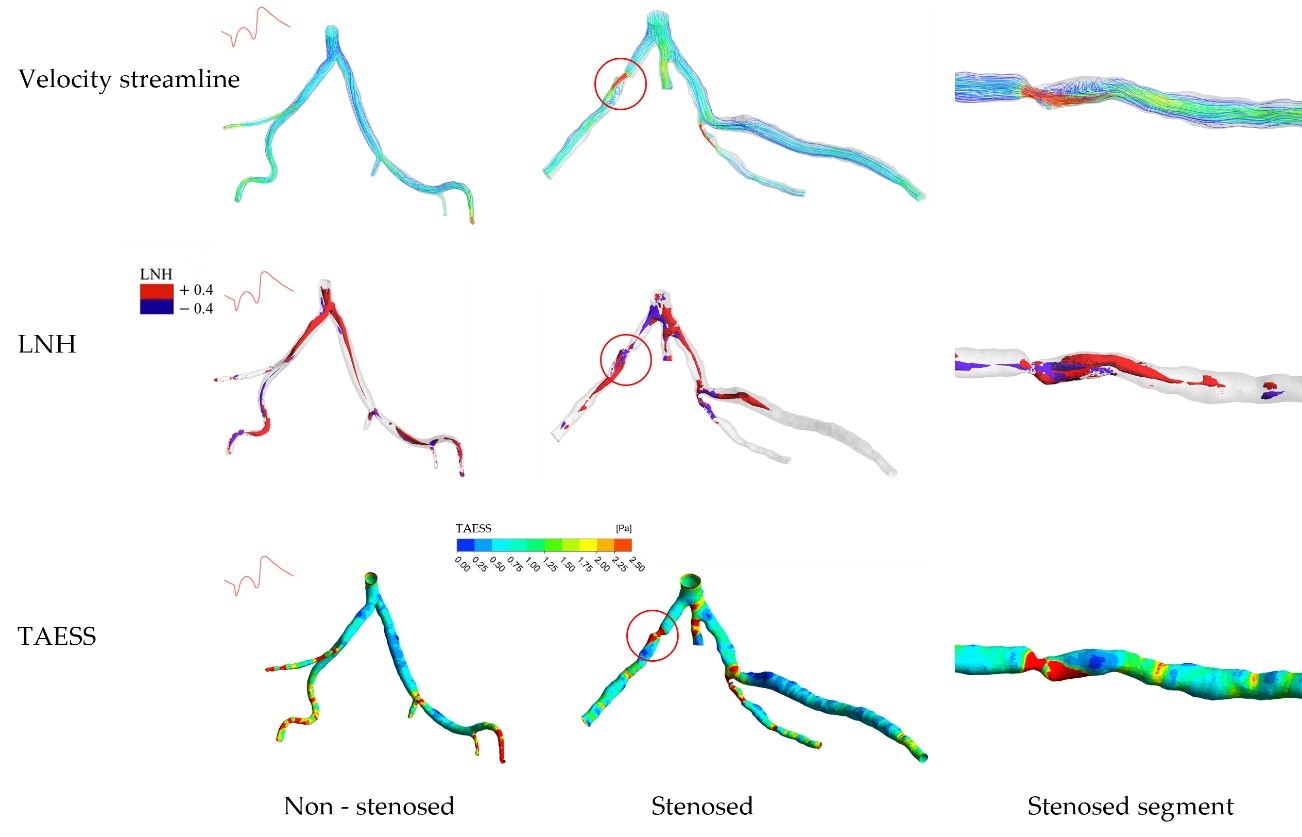}
    \caption{Blood flow characteristics in a patient-specific coronary artery tree: one non-stenosed case (left), one $>70$\% stenosed case (middle), and a zoomed-in view of the stenosed segment (right) highlighted by the red circle. Examples of time-averaged velocity streamlines (top), time-averaged Localised Normalised Helicity (LNH) (middle), and Time-Averaged Wall Shear Stress (TAESS) contour map (bottom) are shown. Disturbed flow patterns were evident in the stenosed regions and downstream, characterised by jet flows and recirculations. More intense and unbalanced HF patterns occurred in the stenosed segments. }
    \label{HF_figure2}
\end{figure}

Comparing the general haemodynamics distributions in artery trees from non-stenosed and stenosed groups, the existence of stenosis affected local haemodynamics, as shown in Figure \ref{HF_figure2}. Disturbed flow was evident in the stenosed regions and downstream of the stenoses. Jet flows developed at the stenosis with high TAESS distributions. Recirculation appeared immediately after the stenosis, where more intense and unbalanced HF patterns were also observed. Both low and high TAESS were downstream of the stenosis (Figure \ref{HF_figure2}). Here, the direct comparison between segments with and without stenosis cannot be conducted since the non-stenosed and stenosed arteries were not from the same patient. Thus, a vessel-specific comparison of the stenosis-free segments from both groups was applied.

The vessel-specific comparison between non-stenosed and stenosed groups revealed a significant difference in HF $h_1$ ($p = 0.0483$) and $h_4$ ($p = 0.0168$) only in the Diagonal, whereas other coronary tree segments had similar HF (Figure \ref{HF_figure2}). No significant difference in $h_2$ and $h_3$ existed between all stenosis-free segments from both groups.

\begin{figure}
    \centering
    \includegraphics[width=1\linewidth]{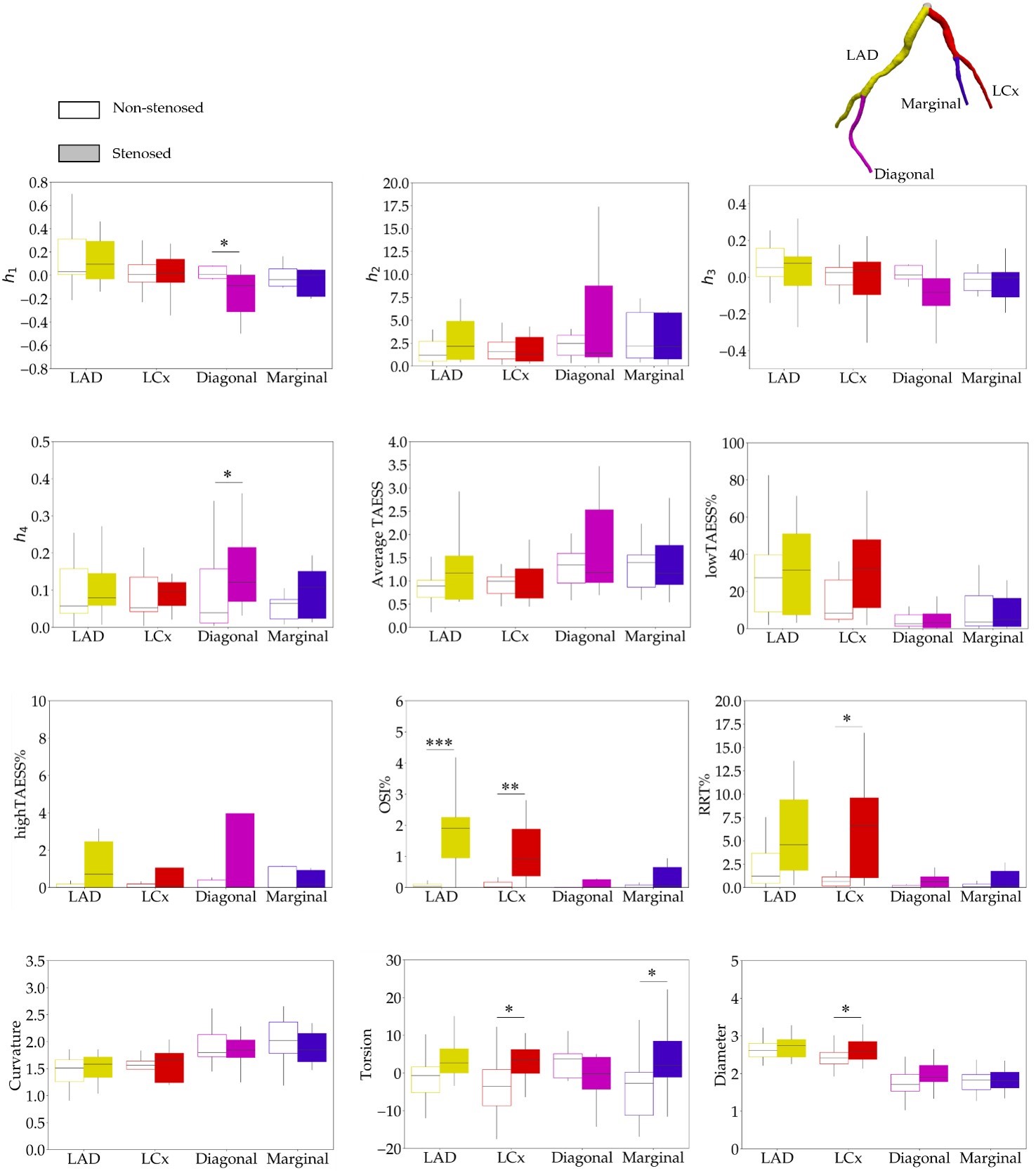}
    \caption{Boxplots illustrating Helical Flow (HF) and near-wall haemodynamic parameters in non-stenosed (unfilled) versus stenosed (filled) segments of the Left Anterior Descending (LAD, yellow), Left Circumflex (LCx, red), Diagonal (pink), and Marginal (blue) coronary arteries. Statistical comparisons were performed using Welch's t-test for normally distributed variables and the Mann-Whitney U-test for non-normally distributed variables, with a significance threshold of $p < 0.05$. Statistically significant differences are denoted by asterisks (* $p < 0.05$, ** $p < 0.01$, *** $p < 0.001$).
}
    \label{HF_figure3}
\end{figure}

No significant differences were found for average TAESS, lowTAESS\% or highTAESS\%. OSI\% was significantly higher in stenosed LAD and LCx cases, whereas RRT\% was significantly higher in LCx (Figure \ref{HF_figure3}, OSI\% LAD: $p < 0.0001$, non-stenosed groups: $0.03 \, [0.01–0.11]$ vs stenosed group: $1.90 \, [0.95–2.25]$; OSI\% LCx: $p = 0.0008$, non-stenosed groups: $0.02 \, [0.00–0.17]$ vs stenosed group: $0.92 \, [0.37–1.87]$; RRT\% LAD: $p = 0.0185$, non-stenosed groups: $1.23 \, [0.44–3.66]$ vs stenosed group: $4.56 \, [1.84–9.40]$; RRT\% LCx: $p = 0.0323$, non-stenosed groups: $0.65 \, [0.17–1.13]$ vs stenosed group: $6.61 \, [1.04–9.59]$).

Considering the geometrical factors (Figure \ref{HF_figure3}), the diameter ($p = 0.0411$) and torsion ($p = 0.0238$) of the LCx artery were significantly larger in stenosed cases compared to non-stenosed cases. Additionally, the Marginal torsion was also greater in stenosed cases ($p = 0.0425$).

\FloatBarrier

\subsection{Correlations between HF-based, ESS-derived and geometrical factors in stenosis-free segments}

When considering the effect of HF in stenosis-free segments in stenosed and non-stenosed trees, in both non-stenosed and stenosed cases, $h_2$ positively correlated with the average TAESS in all the segments ($p < 0.05$), indicating that the higher absolute helicity, $h_2$, is associated with higher absolute TAESS across arterial segments if no disease is present locally, irrespective of disease globally. $h_1$, $h_3$, and $h_4$ showed minor effects on haemodynamics in all arterial segments.

In trees with global stenosis, $h_2$ significantly correlated to lowTAESS\% in all segments if no disease was present locally (LAD: $\rho = -0.86, p = 0.0015$, LCx: $\rho = -0.89, p = 0.0002$, Diagonal: $\rho = -0.80, p = 0.0098$, Marginal: $\rho = -0.94, p < 0.0001$). $h_2$ also correlated with highTAESS\% in LAD, LCx, and Diagonal (LAD: $\rho = 0.85, p = 0.0026$, LCx: $\rho = -0.76, p = 0.0322$, Diagonal: $\rho = -0.84, p = 0.0021$).

However, in trees without any stenosis, no correlations were found between $h_2$ and other haemodynamics in any segments. Since $h_2$ did not correlate with lowTAESS\%, highTAESS\%, OSI\%, and RRT\%, $h_2$ will not influence the lumen area exposed to adverse haemodynamics in non-stenosed cases.

When considering the effect of geometrical factors in stenosis-free segments from stenosed and non-stenosed cases, only the diameter of Diagonal arteries negatively correlated with average TAESS in non-stenosed cases ($\rho = -0.85, p = 0.0046$), and $h_1$ in the stenosed cases ($\rho = -0.75, p = 0.0361$). No other significant correlations were found.

\subsection{Correlations between HF-based, ESS-derived and geometrical factors in stenosed segments }

Within the stenosed arteries, $h_2$ positively correlated with highTAESS\% ($\rho = 0.87, p = 0.0004$), average TAESS ($\rho = 0.96, p < 0.0001$), and negatively correlated with lowTAESS\% ($\rho = -0.89, p = 0.0001$). In pre-stenosis segments, only $h_2$ positively correlated with average TAESS ($\rho = 0.84, p = 0.0444$), while no significant correlations were observed between other HF-based descriptors and ESS-derived factors.

\begin{figure}
    \centering
    \includegraphics[width=1\linewidth]{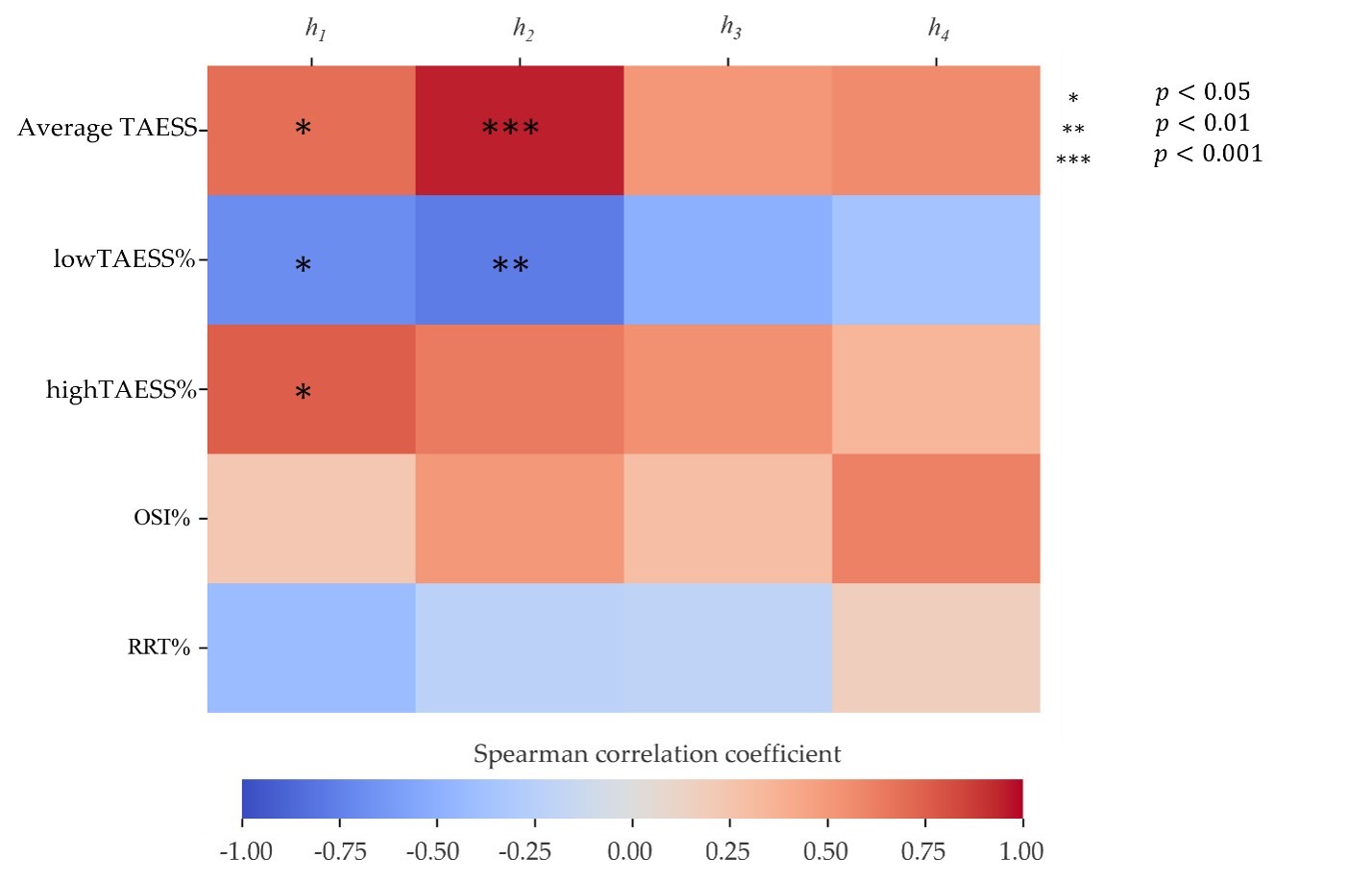}
    \caption{Heatmap showing the Spearman correlation coefficients between haemodynamics and HF-based descriptors in post-stenosis segments. The horizontal colour bar represents the Spearman correlation coefficient scale. Statistically significant correlations are indicated by asterisks. $h_1$ quantifies the helicity intensity considering the rotational directions of the counter-rotating HF, and $h_2$ measures the absolute helicity. Only helicity intensity ($h_1$ and $h_2$) showed significant correlations with haemodynamic descriptors. The signed ($h_3$) and unsigned ($h_4$) helical rotation balance did not show significant correlations with ESS-derived factors.
}
    \label{HF_figure4}
\end{figure}

The presence of stenosis significantly altered the local flow conditions (Figure \ref{HF_figure2}), thereby HF showed significant correlations with both low and high TAWSS. Figure \ref{HF_figure4} illustrates the Spearman correlation coefficients $\rho$ between HF-based and ESS-derived factors in post-stenosis segments. Within and after the stenoses, jet flow induced high HF and high TAESS (Figure \ref{HF_figure2}, right), resulting in strong correlations between $h_2$ and highTAESS\% at the stenosis ($\rho = 0.86, p = 0.0016$), and between $h_1$ and highTAESS\% after the stenosis ($\rho = 0.75, p = 0.0106$). Due to the unbalance of the HF after the stenosis (Figure \ref{HF_figure2}, right), the signed $h_1$ can better quantify the HF intensity, which explains why there is no significant correlation between $h_2$ and highTAWSS\%.

Blood flow recirculations appeared immediately after the stenosis (Figure \ref{HF_figure2}, right), resulting in low TAESS regions at the site of flow recirculation. Both high signed ($h_1$) and unsigned ($h_2$) helicity intensity showed negative correlations with lowTAWSS\%, indicating that high-intensity HF mitigated the flow disturbance after stenosis. Although unbalanced HF was observed in the post-stenosed segment regions (Figure \ref{HF_figure2}, right), $h_3$ and $h_4$, quantifying the HF unbalance, did not show significant correlations (Figure \ref{HF_figure4}).

\newpage
\section{Discussion}

In this work, we investigated the influence of HF on haemodynamics and considered the coronary geometrical effect in 39 patient-specific artery trees using the open-source ASOCA dataset (20 non-stenosed vs 19 stenosed left coronary arteries) \cite{GharleghiASOCA2022,Gharleghi2023ASOCADATA}. Consistent with previous findings \cite{Nisco2019HF,Nisco2020HF}, HF with high absolute intensity ($h_2$) correlated with an increase in TAESS in both non-stenosed and stenosed arteries, thereby reducing the arterial areas exposed to adversely low TAESS. However, we demonstrated for the first time that this also increases arterial areas exposed to adversely high TAESS in stenosed arteries.
Previous studies have reported that stenosis increases local HF intensity \cite{Chiastra2017Healthy_and_diseased} and TAESS \cite{Frattolin2015_disease,Kamangar2017_disease_hyperemic}. However, the plaque can further progress or even rupture around the stenosis, making the high-intensity HF's effect unclear. Our results indicated that high-speed and high-intensity HF in stenosed regions negatively affects the vessel wall, leading to excessively high TAESS. This has been linked to exacerbated plaque vulnerability, leading to destabilisation and rupture \cite{Eshtehardi2017_high_WSS,bajraktari2021_highTAWSS}, thereby increasing the risk of severe cardiac events \cite{Kumar2018highwss}. At the same time, high-intensity HF can still mitigate exposure to low TAESS, which is caused by flow disturbances and low-velocity zones due to recirculations. Consequently, high-intensity HF in stenosed segments demonstrates a double-edged nature. While it reduces luminal areas exposed to adversely low TAESS, it also increases adversely high TAESS in stenosed regions. Therefore, it can be argued that for already diseased cases, more emphasis should be placed on the adverse impact of localised high-intensity HF on existing plaque.
The positive correlation between HF intensity and high TAESS in stenosis-free segments in stenosed groups can partially explain the previous findings that SCAD \cite{Candreva2023_SCAD} and non-obstructive myocardial ischemia \cite{ZebicMihic2023_NonObstructive_CAD,Estrada2022_tortuosity_MI} observed in curved arterial segments, where high HF intensity was observed \cite{Vorobtsova2016tortuosity_BA,Chiastra2017Healthy_and_diseased,Shen2021Secondary_flow}. Both SCAD \cite{Candreva2023_SCAD} and non-obstructive myocardial ischemia \cite{Kumar2018highwss,Candreva2022highwss} have been linked to adversely high TAESS. In fact, SCAD was co-localised with higher TAESS (90th percentile) in extremely curved segments, and vessels with healed SCAD showed lower TAESS peak values \cite{Candreva2023_SCAD}, which indicates the effect of adversely high TAESS on SCAD development. Besides, compared to low TAESS, which is associated with decreased lumen area (constrictive remodelling), adversely high TAESS is related to expansive remodelling with high plaque vulnerability \cite{Samady2011,bajraktari2021_highTAWSS}. Although we did not examine the HF in coronary arteries with SCAD or non-obstructive myocardial ischemia, the association found between high HF intensity and adversely high TAESS can help to explain the apparent contradiction between the reported protective effects of high HF intensity \cite{Nisco2019HF,Nisco2020HF} and the clinical outcomes observed in severely curved arteries \cite{Tuncay2018invivo_tortuosity_CTA,Candreva2023_SCAD,ZebicMihic2023_NonObstructive_CAD,Estrada2022_tortuosity_MI}. This relationship also provides new insight into disease development in severely curved segments by linking geometry-induced high HF intensity with the risk of excessively high TAESS, warranting future studies.
Interestingly, no correlations between HF intensity and high TAESS were found in the non-stenosed group. In normal arteries, HF naturally and effectively regulates flow and maintains TAESS within a healthy physiological range. However, in arteries with complex geometries, such as severe curvature, or after the disease is onset, overly increased HF intensity can lead to adverse haemodynamics, potentially influencing disease progression. This is in line with recent findings, highlighting the different effects of haemodynamic metrics according to disease stage \cite{Zhang2024_plaque_onset_and_progression}. Therefore, it is crucial to focus on the relationship between HF and adversely high TAESS in vessels with existing disease or complex geometries to better understand the role of HF in the development of atherosclerotic coronary artery disease.
This study has some limitations. Although 39 coronary artery trees were considered in this work, which included more patient-specific geometries than previous studies (n = 3), the distributions of measured geometrical factors showed outliers, indicating individual differences. Consequently, no statistically significant correlations were found between geometrical and HF-based factors, which were reported in previous studies \cite{Vorobtsova2016tortuosity_BA,Chiastra2017Healthy_and_diseased,Shen2021Secondary_flow}. Additionally, the use of Bonferroni-Holm correction in our statistical analysis, aimed at minimising the type I error \cite{Abdi2010_Bonferroni}, may have contributed to this lack of correlation, as this correction was not applied in previous studies. In fact, before applying the correction, significant correlations were observed between geometrical factors with haemodynamic descriptors, suggesting a potential effect of geometrical factors. Thus, studies using larger datasets will be beneficial to HF studies, covering individual variations among the population. Therefore, future studies with larger datasets are warranted to account for individual variability and to further explore geometrical effects on HF. Due to the unavailability of \textit{in vivo} flow measurements, a generic boundary setup was adopted as in previous literature. Scaling laws were used to estimate the inflow rate, and a flow split strategy was applied to the outlets \cite{Giessen2011_boundary_conditions}, which were based on empirically modified from Murray’s Law with a robust correlation between flow and diameter at the inlet and the flow split ratio between daughter branches compared to \textit{in vivo} measurements \cite{Giessen2011_boundary_conditions}. Although a lumped parameter model is commonly applied as outflow conditions for stenosed arteries \cite{Taylor2013_lumped_parameter} to capture the fractional flow reserve under the hyperaemic condition \cite{Kim2010_hyperaemic_condition}, our previous study suggested that haemodynamics showed minor differences across stenosis in cases with different severities under the resting condition when using different outflow strategies \cite{Zhang2023outflow,Zhang2024_plaque_onset_and_progression}. Thus, the scaling law for the inflow and flow split strategy at the outlets can effectively serve as boundary conditions in this work since \textit{in vivo} measurements are unavailable.

\section{Conclusion}

In conclusion, our study provides new insights into the relationship between HF and adversely high TAESS by analysing patient-specific non-stenosed and stenosed coronary arteries. HF with higher $h_2$ in stenosed regions correlated with a larger distribution of adversely high TAESS, which is linked to increased plaque vulnerability and a higher risk of cardiac events. Although high-intensity HF can mitigate flow disturbance by increasing adversely low TAESS in both non-stenosed and stenosed coronaries, it can at the same time cause TAESS to exceed the healthy physiological range, especially in stenosed or complex segments (with severe curvature or similar), potentially exaggerating disease processes associated with adversely high luminal shear stress.

\section{Appendix}

\begin{table}[ht]
\centering
\caption{Correlations between geometrical and haemodynamic factors of each segment from healthy cases with initial or adjusted $p < 0.05$.}
\begin{tabularx}{\textwidth}{|l|l|l|>{\centering\arraybackslash}X|>{\centering\arraybackslash}X|>{\centering\arraybackslash}X|}
\hline
\textbf{Segment} & \textbf{Factor 1} & \textbf{Factor 2} & \textbf{Coefficient} & \textbf{Initial $p$-value} & \textbf{Adjusted $p$-value} \\ \hline
LAD       & Torsion   & RRT\%              & -0.46     & 0.0487    & 1.0000 \\ \hline
LCx       & Torsion   & $h_1$              & -0.50     & 0.0361    & 1.0000 \\ \hline
Diagonal  & Torsion   & $h_1$              & 0.65      & 0.0114    & 0.5009 \\ 
          & Torsion   & $h_3$              & 0.56      & 0.0389    & 1.0000 \\ 
          & Torsion   & $h_2$              & -0.62     & 0.0186    & 0.7814 \\ \hline
Marginal  & Diameter  & Average TAESS      & -0.85     & 0.0001    & 0.0046 \\ 
          & Diameter  & highTAESS\%        & -0.62     & 0.0169    & 0.7255 \\ 
          & Diameter  & lowTAESS\%         & 0.70      & 0.0039    & 0.1804 \\ 
          & Diameter  & Average TAESS      & -0.68     & 0.0054    & 0.2437 \\ 
          & Diameter  & RRT\%              & 0.61      & 0.0155    & 0.6367 \\ 
          & Curvature & lowTAESS\%         & -0.60     & 0.0172    & 0.6880 \\ \hline
\end{tabularx}
\label{table:correlations_healthy}
\end{table}

\begin{table}[ht]
\centering
\caption{Correlations between geometrical and haemodynamic factors of each diseased-free segment from diseased cases with initial or adjusted $p < 0.05$.}
\begin{tabularx}{\textwidth}{|l|l|l|>{\centering\arraybackslash}X|>{\centering\arraybackslash}X|>{\centering\arraybackslash}X|}
\hline
\textbf{Segment} & \textbf{Factor 1} & \textbf{Factor 2} & \textbf{Coefficient} & \textbf{Initial $p$-value} & \textbf{Adjusted $p$-value} \\ \hline
Diagonal & Diameter  & $h_1$          & -0.75      & 0.0008    & 0.0361 \\ 
         & Diameter  & $h_3$          & -0.55      & 0.0273    & 1.0000 \\ 
         & Curvature & $h_1$          & 0.57       & 0.0218    & 0.9162 \\ 
         & Curvature & $h_3$          & 0.57       & 0.0218    & 0.9162 \\ \hline
Marginal & Diameter  & RRT\%          & 0.62       & 0.0179    & 0.7716 \\ \hline
\end{tabularx}
\label{table:correlations_diseased}
\end{table}

\clearpage
\section{References}

\end{document}